\documentstyle[epsfig,12pt,fleqn]{article}

\def\theequation{\thesection.\arabic{equation}}

\def\be{\begin{equation}}
\def\ee{\end{equation}}
\def\bea{\begin{eqnarray}}
\def\eea{\end{eqnarray}}

\def\fnot{\hspace{-0.2cm}/}
\begin{document}
\begin{titlepage}
\begin{center}
{\Large\bf Rare decays $\bar{B}\to X_{s(d)}\gamma$ at the NLO}
\end{center}
\vspace{3cm}
\begin{center}
{\large\bf{
H. M. Asatrian,  H. H. Asatryan, A. Hovhannisyan}}\\
\vspace{1cm}
\it{Yerevan Physics Institute, 2 Alikhanyan Br., 375036, Yerevan, Armenia}\\
\vspace{2.5cm}
\bf{ Abstract}
\end{center}
We present an independent calculation of ${\cal{O}}(\alpha_s)$ matrix elements of the QCD
penguin operators for the decays $\bar{B}\to X_{s(d)}\gamma$. The Mellin-Barnes
representation technique is applied for the systematic numerical evaluation of involved two-loop
diagrams. The numerical effect of the calculated contributions is confirmed to be of
order of $1\%$. We also present an updated numerical analysis for $\bar{B}\to X_{s(d)}\gamma$ decay rates.
\end{titlepage}

\section{Introduction}
The rare B decays play a special role in the phenomenology of weak decays. They are loop-induced
in the standard model (SM) and thus are strongly affected by many contributions
from the new sources of flavor violation
in various extensions of the SM. The inclusive rare transitions are of particular interest as they are
well approximated by the underlying partonic transitions. The corrections to this perturbative 
picture are calculable within the framework of Heavy Quark Effective Theory (HQET).

Among the rare B-decays the inclusive $\bar{B}\to X_s\gamma$ transition is the most interesting one.
Since the first observation of the decay
by the CLEO \cite{Alam:1995aw} new measurements have been
reported by ALEPH, BELLE and BABAR \cite{Barate:1998vz}-
\cite{Abe:2001hk}. The CLEO still has the best
determination \cite{Chen:2001fj}:
\be
{\cal{B}}(\bar{B}\to X_s\gamma)=(3.21\pm 0.43\pm 0.27_{-0.10}^{+0.18})\times 10^{-4}.
\ee
while the current world average is \cite{Stone:2003pc}
\be
{\cal{B}}(\bar{B}\to X_s\gamma)=(3.40\pm 0.39)\times 10^{-4}.
\ee
The CKM-suppressed counterpart $\bar{B}\to X_d\gamma$ decay is also promising.
However, no experimental observations of the $b\to d\gamma$ mediated transitions
have been made so far. The upper experimental limits for the exclusive decays
$B\to \rho^{\pm(0)}\gamma, B\to \omega\gamma$ \cite{Aubert:2003me} are higher than the
corresponding theoretical predictions roughly by a factor of 2.

As for the theoretical side, calculation of the NLO perturbative QCD corrections for the decay
$\bar{B}\to X_s\gamma$ has been  completed in \cite{Buras:xp}-\cite{Buras:2002tp}. 
These corrections lead
to dramatic reduction of the renormalization scale dependence of the leading order (LO)
results bringing the uncertainty in theoretical predictions for the decay rate to the level
of 10\% and opening the door for a rigorous comparison with experimental data. 
The leading order long distance corrections
within the framework of HQET have been computed in \cite{Falk:1994dh}-\cite{Buchalla:1998ky}.
Also the leading electroweak corrections are known \cite{Gambino:2000fz}.
The comparison of the current theoretical predictions with the available experimental data already
presents strong constraints on the extensions of the SM (see e.g. \cite{Gambino:2001ew} 
for a recent analysis).
These analyses have been
extended to the case of $\bar{B}\to X_d\gamma$ decay in \cite{Ali:1998rr}-\cite{Asatryan:2000kt}. 

Only about two years
ago the last missing ingredients of the full NLO result- ${\cal O}(\alpha_s)$ matrix elements of the
QCD-penguin operators, have been calculated in \cite{Buras:2002tp}.
The calculation is complicated due to large number of contributing 
non-trivial two-loop diagrams, so
a verifying re-calculation is desirable.
In this paper we present an independent computation of these matrix elements.
We apply the Mellin-Barnes representation
technique for the systematic numerical calculation of the involved two-loop diagrams. This is in
contrast to the method of \cite{Buras:2002tp} where the results are expressed in
terms of a few integrals to be evaluated numerically. We confirm their numerical results and
extend the analysis to the case of $\bar{B}\to X_d\gamma$ decay (see also \cite{Hurth:2003dk}
for a very recent similar analysis).

The paper is organized as follows. In Section 2 we present the effective Hamiltonians for the
decays $\bar{B}\to X_{s(d)}\gamma$ and collect the final results for the matrix
elements of the dimension-6 operators. Section 3 is devoted to the details of
our calculation method. In Section 4 we give an update of numerical predictions
for the branching fractions for $\bar{B}\to X_{s(d)}\gamma$. Some of the useful formulae
are collected in appendices. 

\section{The decay rates for $\bar{B}\to X_{s(d)}\gamma$}
\setcounter{equation}{0}

The effective Hamiltonian for the decay $b\to s\gamma$ (and
the associated bremsstrahlung process $b\to s\gamma g$) reads
\begin{equation}\label{heff}
{\mathcal H}_{eff}(b\rightarrow s\gamma)=
-\frac {4 G_F}{\sqrt2}\left(
\lambda_t^{(s)}\sum_{i=1}^{8} C_i O_i
-\lambda_u^{(s)}\sum_{i=1}^{2}
C_i \left( O_i^u- O_i\right)\right)
\end{equation}
where the CKM factors are given by
\be
\lambda_t^{(s)}\equiv V_{ts}^{*}V_{tb}=-A\lambda^2\left(1-\frac{A^2\lambda^4}{2}\right)
\left(1-\frac{\lambda^2}{2}+\lambda^2(\rho-i\eta)\right),
\,\,\, \lambda_u^{(s)}\equiv V_{us}^{*}V_{ub}=A\lambda^4(\rho-i\eta).
\ee
The Wilson coefficients $C(\mu)$ are known to ${\cal O}(\alpha_s)$ precision.
We refer to \cite{Chetyrkin:1996vx} and references therein for the explicit formulae.
The dimension six effective operators can be chosen as
\be\label{operators}
\begin{array}{ll}
O^u_1=(\bar s_L \gamma_\mu T^a u_L)(\bar u_L \gamma^\mu T^a b_L),&
O^u_2=(\bar s_L \gamma_\mu u_L)(\bar u_L \gamma^\mu b_L),\vspace{0.2cm}\\
O_1=(\bar s_L \gamma_\mu T^a c_L)(\bar c_L \gamma^\mu T^a b_L),&
O_2=(\bar s_L \gamma_\mu c_L)(\bar c_L \gamma^\mu b_L),\vspace{0.2cm}\\
O_3=(\bar s_L \gamma_\mu b_L)\sum_q (\bar q \gamma^\mu q),&
O_4=(\bar s_L \gamma_\mu T^a b_L)\sum_q (\bar q \gamma^\mu T^a q),\vspace{0.2cm}\\
O_5=(\bar s_L \gamma_\mu\gamma_\nu\gamma_\rho b_L)
    \sum_q (\bar q \gamma^\mu\gamma^\nu\gamma^\rho q),&
O_6=(\bar s_L \gamma_\mu\gamma_\nu\gamma_\rho T^a b_L)
    \sum_q (\bar q \gamma^\mu\gamma^\nu\gamma^\rho T^a q),\vspace{0.2cm}\\
O_7=\frac {e}{16\pi^2}m_b(\bar s_L \sigma^{\mu\nu}b_R)F_{\mu\nu},&
O_8=\frac {e}{16\pi^2}m_b(\bar s_L \sigma^{\mu\nu}b_R)F_{\mu\nu},
\end{array}
\ee
The corresponding formulae for the decays $b\to d\gamma(g)$ can be obtained
from Eqs.(\ref{heff}),(\ref{operators}) with the obvious replacement $s\to d$.
The relevant CKM factors read
\be
\lambda_t^{(d)}\equiv V_{td}^{*}V_{tb}=A\lambda^3
\left(1-\frac{A^2\lambda^4}{2}\right)\left(1-\bar{\rho}+i\bar{\eta}\right),
\,\,\, \lambda_u^{(d)}\equiv V_{ud}^{*}V_{ub}=A\lambda^3(\bar{\rho}-i\bar{\eta}).
\ee
Taking into account that $\lambda$ is small ($\lambda\simeq 0.22$) one can safely neglect
the terms proportional to $\lambda_u^{(s)}$ in the case of $b\to s\gamma$. In the case of
$b\to d\gamma$ however these terms are important in particular being 
responsible for the generation of a large CP-asymmetry. We will keep those terms both in 
analytical and numerical formulae.

The decay widths for the partonic transitions can be expressed as
\begin{eqnarray}
\nonumber
\Gamma[b&\to& s(d)\gamma]^{E_\gamma>E_0\equiv(1-\delta)m_b/2}=\\
&=&\frac{G_F^2\alpha_{em}}{32\pi^4}\left|\lambda_t^{s(d)}\right|^2
m_{b,pole}^3 m^2_{b,\overline{MS}}(m_b)
\left(\left|D^{s(d)}+\epsilon_{{\rm ew}}\right|^2+A^{s(d)}\right),
\label{cutoff}
\end{eqnarray}
where we have separated the virtual and bremsstrahlung corrections. At the NLO we have
\begin{eqnarray}\nonumber
D^{s(d)}&=&C_7^{(0)eff}(\mu_b)+
\frac{\alpha_s(\mu_b)}{4\pi}
\left(C_7^{(1)eff}(\mu_b)+\sum_{i=1}^8
C_i^{(0)eff}(\mu_b)\left[r_i+\gamma_{i7}^{(0)eff}\ln\frac{m_b}{\mu_b}\right]\right)\\
&-&\frac{\lambda_u^{s(d)}}
{\lambda_t^{s(d)}}\frac{\alpha_s}{4\pi}{\sum_{i.j=1}^2}C_i^{(0)eff}(\mu_b)
[r_i^u-r_i],
\\
\nonumber
A^{s(d)}&=&\frac{\alpha_s}{\pi}\left\{\sum\limits_{i,j=1}^{8}C_iC_jf_{ij}^{cc}
-2{\rm Re}\left[\frac{\lambda_u^{s(d)}\lambda_t^{s(d)}}{|\lambda_t^{s(d)}|^2}
\sum_{i=1}^{8}\sum_{j=1}^{2}C_iC_j(f_{ij}^{uc}-f_{ij}^{cc})\right]\right.
\\
&+&\left.\left|\frac{\lambda_u^{s(d)}}{\lambda_t^{s(d)}}\right|^2\sum_{i,j=1}^{2}C_iC_j
\left[f_{ij}^{uu}+f_{ij}^{cc}-2{\rm Re}f_{ij}^{uc}\right]\right\}.
\end{eqnarray}
The formulae for the bremsstrahlung coefficients $f_{ij}$ are presented in Appendix B. 
$\epsilon_{{\rm ew}}$ represents the leading electroweak 
correction that can be extracted from \cite{Gambino:2000fz}.

The quantities $r_i$ are given by
{\small{\be\label{ri}
\begin{array}{ll}
r_1=-\frac{1}{6}r_2,\hspace{3cm}&
r_2=-\frac{1666}{243}+2\left(a(z)+b(z)\right)-\frac{80}{81}i\pi,
\vspace{0.2cm}\\
r_3=10.059+\frac{28}{81}i\pi,&
r_4=-1.017-\frac{110}{243}i\pi+2b(z),
\vspace{0.2cm}\\
r_5=185.841+\frac{448}{81}i\pi,&
r_6=-8.148-\frac{2480}{243}i\pi+12a(z)+20b(z),
\vspace{0.2cm}\\
r_7=-\frac{10}{3}-\frac{8\pi^2}{9},&
r_8=\frac{44}{9}-\frac{8}{27}\pi^2+\frac{8}{9}i\pi,
\end{array}
\ee}}
\hspace{-0.6cm}and $r_i^u=\displaystyle\lim_{z\to 0}r_i (i=1,2)$. We will
present the calculation of the matrix elements $r_3,..., r_6$ in some detail
in Section 3.
In Eq. (\ref{ri}) we have used the functions $a(z), b(z)$ 
defined in \cite{Buras:2002tp}
to encode $z$-dependent part of the matrix elements of the current-current operators
$O_1$ and $O_2$:
{\small{\begin{eqnarray}
a(z)&=&\frac{16}{9}\left\{\left(\frac{5}{2}-\frac{1}{3}\pi^2-3\zeta(3)+\left(\frac{5}{2}-\frac{3}{4}\pi^2\right)L
+\frac{1}{4}L^2+\frac{1}{12}L^3\right)\,z\right.\\
\nonumber
&&+\left(\frac{7}{4}+\frac{2}{3}\pi^2-\frac{1}{2}\pi^2L-\frac{1}{4}L^2+\frac{1}{12}L^3\right)\,z^2
+\left(-\frac{7}{6}-\frac{1}{4}\pi^2+2L-\frac{3}{4}L^2\right)\,z^3\\
\nonumber
&&+\left(\frac{457}{216}-\frac{5}{18}\pi^2-\frac{1}{72}L-\frac{5}{6}L^2\right)\,z^4+
\left(\frac{35101}{8640}-\frac{35}{72}\pi^2-\frac{185}{144}L-\frac{35}{24}L^2\right)\,z^5\\
\nonumber&&
+\left(\frac{67801}{8000}-\frac{21}{20}\pi^2-\frac{3303}{800}L-\frac{63}{20}L^2\right)\,z^6+i\pi
\left[\left(2-\frac{1}{6}\pi^2+\frac{1}{2}L+\frac{1}{2}L^2\right)\,z\right.\\
\nonumber &&
\left.\left.
+\left(\frac{1}{2}-\frac{1}{6}\pi^2-L+\frac{1}{2}L^2\right)\,z^2+z^3+\frac{5}{9}z^4+\frac{49}{72}z^5
+\frac{231}{200}\,z^6)\right]\right\}+O(z^7),
\\
b(z)&=&-\frac{8}{9}\left\{\left(-3+\frac{1}{6}\pi^2-L\right)\,z-\frac{2}{3}\pi^2z^{3/2}
+\left(\frac{1}{2}+\pi^2-2L-\frac{1}{2}L^2\right)\,z^2\right.\\
\nonumber &&+\left(-\frac{25}{12}-\frac{1}{9}\pi^2
-\frac{19}{18}L+2L^2\right)\,z^3+\left(-\frac{1376}{225}+\frac{137}{30}L+2L^2+\frac{2}{3}\pi^2\right)\,z^4\\
\nonumber&&
+\left(-\frac{131317}{11760}+\frac{887}{84}L+5L^2+\frac{5}{3}\pi^2\right)\,z^5
+\left(-\frac{2807617}{97200}+\frac{16597}{540}L+14L^2+\frac{14}{3}\pi^2\right)\,z^6\\
\nonumber &&
\left.
+i{\pi}\left[-z+(1-2L)\,z^2+(-\frac{10}{9}+\frac{4}{3}L)\,z^3+z^4+\frac{2}{3}z^5+\frac{7}{9}z^6\right]\right\}
+O(z^7),
\end{eqnarray}}}
where $L=\ln\,z$ \footnote{Originally these functions have been calculated in \cite{Greub:1996tg} 
with precision of $O(z^3)$ and later extended to include terms 
up to $z^6$ in \cite{Buras:2001mq}. We have verified these expressions
using the method of \cite{Greub:1996tg} to include terms 
${\mathcal O}(z^4)$, ${\mathcal O}(z^5)$, and ${\mathcal O}(z^6)$.}.

Following \cite{Gambino:2001ew} we use the following formula for the branching ratios:
\be
\label{branching}
{\mathcal B}(\bar{B}\to X_{s(d)}\gamma)={\mathcal B}(B\to X_ce\bar{\nu})^{exp}
\frac{\Gamma(\bar{B}\to X_{s(d)}\gamma)}{C\left|V_{cb}/V_{ub}\right|^2\,\Gamma(B\to X_ue\bar{\nu})}
\ee
The essential reason for introducing (\ref{branching}) is that the factor 
$$C=\left|\frac{V_{ub}}{V_{cb}}\right|^2\frac{\Gamma(B\to X_ce\bar{\nu})}{\Gamma(B\to X_ue\bar{\nu})}$$
can be predicted with relatively small uncertainty using 
the $\Upsilon$ expansion \cite{Hoang:1998hm}. 
We use $C=0.575\times(1\pm 0.03)$ as estimated in \cite{Gambino:2001ew}. As it follows from
Eq. (\ref{branching}) ${\cal{O}}(\alpha_s)$ corrections for the inclusive decay $B\to X_ue\bar{\nu}$
have to be included in our analysis.

When doing the transition to B-meson decay rates in Eq. (\ref{branching}) from the corresponding
partonic decay rates one
has to include the leading $1/m_b^2$ \cite{Falk:1994dh}  and $1/m_c^2$ corrections \cite{Buchalla:1998ky}.
Note that in the case of $\bar{B}\to X_d\gamma$ additional long-distance contributions due to intermediate
$u$-quark loops can have a sizeable effect. These contributions can not be reliably 
predicted but the model calculations
indicate that they are not very large \cite{Ricciardi:1995jh},\cite{Deshpande:1994cn}. 
In the present paper we will ignore them. 

\section{Calculation of the virtual ${\mathcal{O}}\left(\alpha_s\right)$ corrections for the
operators $O_3,...,O_6$.}
\setcounter{equation}{0}
\subsection{Organization of the calculation}

\begin{figure}[t]
\centerline{
\epsfysize=2.1in
\epsffile{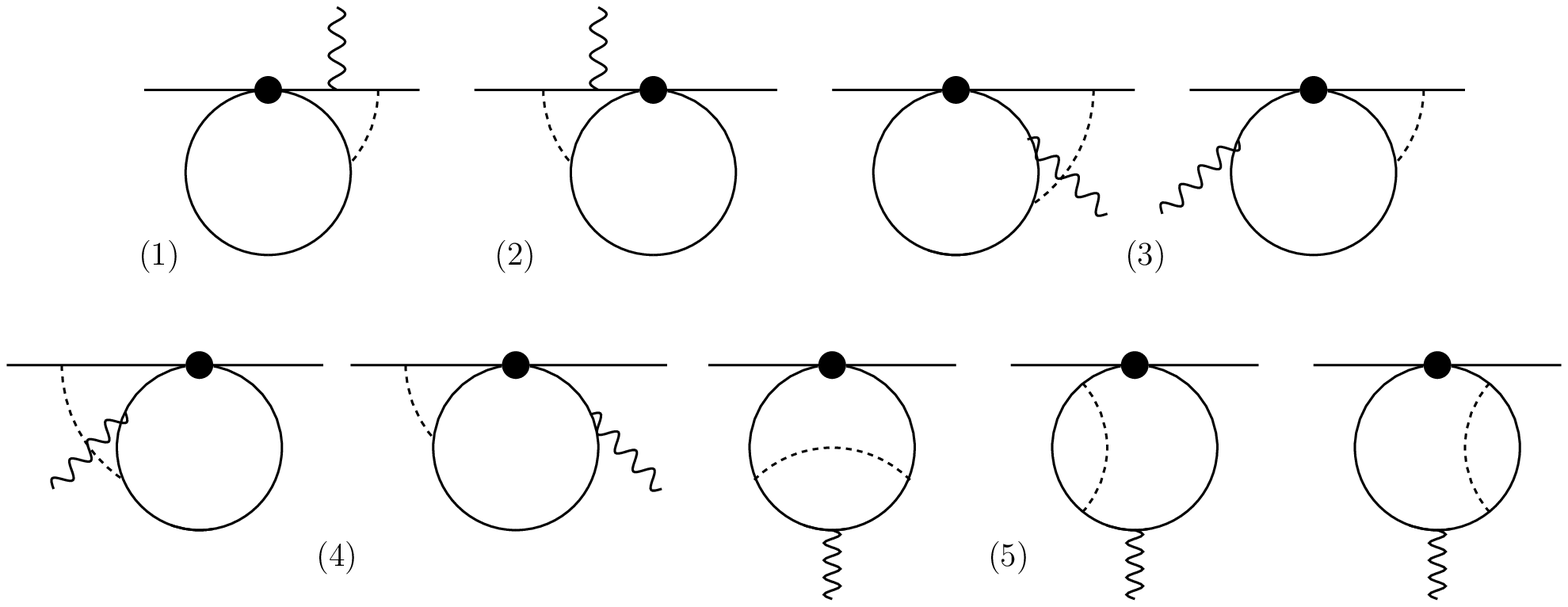}
}
\caption[]{ The diagrams contributing to the matrix elements of the operators $O_3,...,O_6$.
\label{fig:allDiagrams}}
\end{figure}

We start describing the virtual ${\cal{O}}(\alpha_{s})$
contributions to the matrix elements of the operators $O_3,...,O_6$ . 
The relevant Feynman diagrams are depicted in Fig.\ref{fig:allDiagrams}. 
Note that the full set of contributing
diagrams is larger, however, the calculation of 
the other diagrams can be avoided if one uses the gauge
invariance of the result. Furthermore, the contributions of some of the diagrams are already included
when one uses the effective Wilson coefficient $C_7^{eff}$ instead of $C_7$.

\begin{figure}[t]
\centerline{
\epsfxsize=3.0in
\epsffile{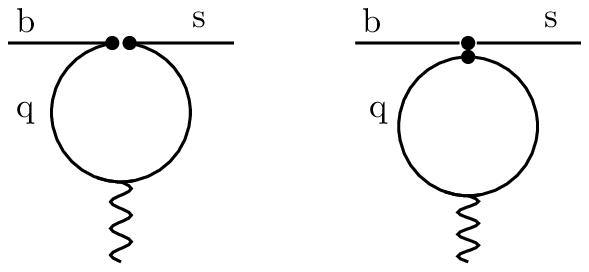}
}
\caption[]{{\bf{ Two types of operator insertions contributing to $b\to s(d)\gamma$.}}
\label{fig:types}}
\end{figure}

Note that for each of the diagrams
there are two possible types of operator insertions as illustrated in Fig.\ref{fig:types}.
Apparently for type I insertions only $b$ and $s$ quark loops contribute while all five
($u$, $d$, $s$, $c$ and $b$) quarks can give non-zero type II contributions. In our calculations we set the
masses of the light quarks $u$, $d$, $s$ to zero. 

Our calculation extends the method described in \cite{Greub:1996tg}. 
While for the light and $c$ quark contributions it is
straightforward, for the evaluations of the diagrams with $b$ quark in the loop we use a different tactics to get
numerical results. As we think that our method can be useful in other similar calculations we describe it in some
detail in subsection \ref{calcMethod}.

For a better organization of our computation we will use the following decomposition:
\begin{eqnarray}
\nonumber && O_3=(\bar{s} \gamma_{\mu}Lb)\sum_q (\bar{q}\gamma^\mu Lq)+
                             (\bar{s} \gamma_{\mu}Lb)\sum_q (\bar{q}\gamma^\mu Rq)
                             \equiv O_{3L}+O_{3R},\\
\nonumber && O_4=(\bar{s}\gamma_{\mu}LT^ab)\sum_q (\bar{q}\gamma^{\mu}LT^a q)+
                            (\bar{s}\gamma_{\mu}LT^ab)\sum_q (\bar{q}\gamma^{\mu}RT^a q)
                            \equiv O_{4L}+O_{4R}
\end{eqnarray}
and similarly for the operators $O_5, O_6$.

Following the ref. \cite{Greub:1996tg}, as a first step in the calculation of the two-loop diagrams 
in Fig.\ref{fig:allDiagrams} we evaluate the building blocks from Fig.\ref{fig:buildingBlocks}. 
We present the full expressions of the one-loop building blocks in Appendix A. As we are going to 
illustrate the method of our calculation in the next subsection we will give here the
expression for the contribution of $O_{3L}$ with $b$-quark loop to the building block $I_{\beta}$:
\bea\label{O3bb}
I_{\beta(3)}^{b(I)}(L)&=&  -\frac{g_s}{4\pi^2}
  \Gamma(\epsilon)(1-\epsilon)  \mu^{2\epsilon}e^{\gamma_{E}\epsilon}e^{i\pi\epsilon}
  \int\limits_{0}^{1}\frac{dx{[x(1-x)]}^{1-\epsilon}}{\Delta^{\epsilon}}
  (r_{\beta}r\hspace{-0.2cm}/-r^2\gamma_{\beta})L\,\frac{\lambda}{2}
\eea
where 
\be
\Delta=r^2-{m_q^2}/{x(1-x)}+i\delta.
\ee

\subsection{The method of the calculation\label{calcMethod}}

We will illustrate our method of calculation on the example of diagram 
(1) from Fig.\ref{fig:allDiagrams} for the type I
insertion of the operator $O_{3L}$. Using (\ref{O3bb})
and applying the standard technique of the Feynman parameterization
for the integration over the second 4-momentum we arrive to
\footnote{The details of the calculation of the same diagram with a charm-quark loop for the operator $O_2$ 
are given in \cite{Greub:1996tg}.}
\begin{eqnarray}
\label{example1}
\nonumber
M(1c)&=&\frac{eQ_dg_s^2C_F}{64\pi^4}\Gamma(2\epsilon)\exp^{2i\pi\epsilon}
\int dxdudvdy
[x(1-x)]^{1-\epsilon}y^{\epsilon-1}(1-v)^{\epsilon}v\times\\
&&\bar{u}(p')\left[P_1\frac{\hat{C}}{\hat{C}^{2\epsilon}}+
P_2\frac{1}{\hat{C}^{2\epsilon}}+P_3\frac{1}
{\hat{C}^{1+2\epsilon}}\right]u(p)
\end{eqnarray}
where the integration over the Feynman parameters $x,y,u,v$ is 
over the interval [0;1] and  $\hat{C}$ is given by
\begin{equation}
\label{denom}
\hat{C}=m_b^2v(1-v)u-m_q^2\frac{(1-v)y}{x(1-x)}+i\delta
\end{equation}
The strategy in \cite{Greub:1996tg} (as well as in the similar calculations \cite{Greub:2001sy},
\cite{Asatrian:2001de})
is then to use the Mellin-Barnes representation for the integrals in eq.~(\ref{example1}) 
which is given by 
\begin{equation}
\label{melin}
 \frac{1}{(k^2-M^2)^\lambda}=\frac{1}{(k^2)^\lambda}\frac{1}{\Gamma(\lambda)}\frac{1}{2\pi
i}\int_\gamma ds (-M^2/k^2)^s\Gamma(-s)\Gamma(\lambda+s)
\end{equation}
where $\lambda>0$ and the integration path goes parallel to the imaginary axis in the complex
$s$-plane hitting the real axis somewhere between $-\lambda$ and $0$. In \cite{Greub:1996tg} the
formula (\ref{melin}) is applied to the denominators in (\ref{example1}) using the identification
\begin{equation}
k^2 \leftrightarrow m_b^2v(1-v)u\,\ ;\,\
M^2 \leftrightarrow  m_q^2\frac{(1-v)y}{x(1-x)}
\end{equation}
Then the integrals over the Feynman parameters are expressed in terms of Euler's $\Gamma$-functions.
To evaluate the integral over the Mellin-Barnes parameter $s$ the integration path is closed in the right
$s$-half-plane. When $m_q^2/m_b^2<1/4$ the integral over the half-circle vanishes so using the residue theorem
the integral is expressed as the sum over residues in the integration contour. This in turn leads to a
natural expansion in the (small) parameter $m_q^2/m_b^2$.

This procedure apparently won't work for the case $q=b$. The absence of a small expansion
parameter seems to make the described technique unapplicable in this case. However we note that
the appearance of the factor $x(1-x)\,\,(0\le x\le 1)$ in eq. (\ref{denom})
plays of a role of a suppression factor for the second term and it turns out that the
'opposite' identification
\begin{equation}
M^2 \leftrightarrow  m_b^2v(1-v)u,\,\,\,\,\,\,
k^2\leftrightarrow m_q^2\frac{(1-v)y}{x(1-x)}
\end{equation}
does a better job in the case $q=b$. Just as in \cite{Greub:1996tg} then we first take the integrals
over the Feynman parameters which after appropriate substitutions can be brought into the following forms:
\begin{eqnarray}
\nonumber
\int_0^1dww^p,\,\,\,\,\,\, \int_0^1dww^p(1-w)^q=\frac{\Gamma(p+1)\Gamma(q+1)}{\Gamma(p+q+2)}
\end{eqnarray}
Then to perform the integration over $s$ we can close the integration path in the right half-plane and
use the residue theorem. The relation $x(1-x)\le 1/4$ guarantees then that the integral over the half-circle
vanishes.

So the integrals will be reduced to sums over the residues of $\Gamma$ function that reside on the real
axis (in complex $s$-plane):
\bea
s=n,\,\,\, n-\epsilon,\,\,\, n-2\epsilon \,\,\,\, (n=0,1,2,...),\,\,\,\,\,s=m/2-2\epsilon,\,\,\,\,(m=3,5,...)
\label{poles}
\eea
We note again that though there is no real expansion parameter, the terms with higher values of
$n$ (or $m$) are suppressed because of factor $x(1-x)$. 
So taking into account only first several values of $n$ makes
a good numerical approximation of the final result.

It is easy to check the validity of our procedure for the diagram (1) from Fig. \ref{fig:allDiagrams}.
as it can be calculated analytically (see formula (5.6) of ref. \cite{Buras:2002tp})). Summing over the
residues at $n=0,..,6$ (see Eq.(\ref{poles})) we will get real good approximation
for the final result with the precision of $10^{-6}$. We didn't observe such a fast convergence for the other
diagrams, however the sum over the several first values of $n$ in all of the cases gives a good approximation
for the involved integrals. For each of the diagrams we will keep as many poles as it is necessary for
getting a precision of at least $10^{-3}$. 

\subsection{Results of the calculation}
Using the method described above we are able to calculate the unrenormalized
matrix element of the operator $O_3$.
We find
\begin{eqnarray}
M_3&=&\left ({\frac {248}{81}}\,{\frac
{1}{\epsilon}}\left ({\frac {m_b}{\mu}}\right
)^{-4\epsilon}+9.357
+{\frac {100}{81}}\,i\pi\right )
 \times\frac{\alpha_s}{4\pi}\langle s\gamma |O_7|b\rangle_{tree}
\end{eqnarray}
Looking at the expressions for the building blocks  it is easy to see that
the contributions from the operators can be expressed in terms of corresponding contributions
from $O_3$ with multiplicative factors that can be extracted from the formulae in Appendix A.
The diagrams from Fig. \ref{fig:allDiagrams}.5 that do not contain the building blocks are easy
to calculate without any numerical approximations. The results for the unrenormalized matrix elements
of the remaining operators are
\begin{eqnarray}
\nonumber M_4&=& \left( -{\frac {46}{243\epsilon}}\left ({\frac
{m_b}{\mu}}\right )^{-4\epsilon}- 0.899 +2\,{\it b(z)}-
\frac{146}{243}i\pi\right)
\times\frac{\alpha_s}{4\pi }\langle s\gamma
|O_7|b\rangle_{tree}\\
\nonumber M_5&=&\left({\frac {5552}{81\epsilon}}\left ({\frac
{m_b}{\mu}}\right )^{-4\epsilon} +68.688 +{\frac {1888}{81}}\,i\pi\right)
\times\frac{\alpha_s}{4\pi }\langle s\gamma
|O_7|b\rangle_{tree}\\
\nonumber M_6&=&\left(-\frac{4588}{243\epsilon}\left ({\frac
{m_b}{\mu}}\right )^{-4\epsilon}-25.956+12 a(z)+20 b(z)-\frac{3200}{243}\,i\pi\right)
\times\frac{\alpha_s}{4\pi }\langle s\gamma
|O_7|b\rangle_{tree}
\end{eqnarray}
where
\be
\langle s\gamma|O_7|b\rangle_{tree}=\frac{em_b}{8\pi^2}\bar{u}(p')
\epsilon\hspace{-0.2cm}/q\hspace{-0.2cm}/Ru(p).
\ee
These results are in a perfect agreement with those from \cite{Buras:2002tp}.

The renormalization procedure is described in \cite{Buras:2002tp}. As we follow
the same scheme we don't go into details. The final results for the renormalized
matrix elements are given in Eq. (\ref{ri}).

\section{Numerical results}
\setcounter{equation}{0}

In this section we will present the numerical results for the branching ratios of the decays
$\bar{B}\to X_{s(d)}\gamma$. In our numerical analysis we will use the following values taken
from \cite{PDG}:
\be
M_W=80.425\,{\rm GeV},\,\,\, M_Z=91.1876\,{\rm GeV},\,\,\,m_t^{pole}=174.3\pm 5.1\,{\rm GeV},\,\,\,
\alpha_{em}=1/137.036.
\ee
For the semileptonic branching ratio we will use the averaged value 
${\mathcal B}(B\to X_c\ell^+\bar{\nu}_{\ell})^{exp}=(10.90\pm0.23)\%$ from \cite{SLAC}.
We use two-loop expression for the strong coupling constant $\alpha_s(\mu)$ with
$\alpha_s(M_z)=0.1172\pm 0.002$ \cite{PDG}.
For the Wolfenstein parameters of the CKM matrix we will use \cite{Battaglia:2003in}
\be
\rho=0.162\pm 0.046,\;\;\eta=0.347\pm 0.027,\;\;A=0.83\pm 0.02,\;\;\lambda=0.2240\pm0.0036
\ee
\begin{figure}[t]
\epsfxsize=3.5in
\begin{center}
\epsffile{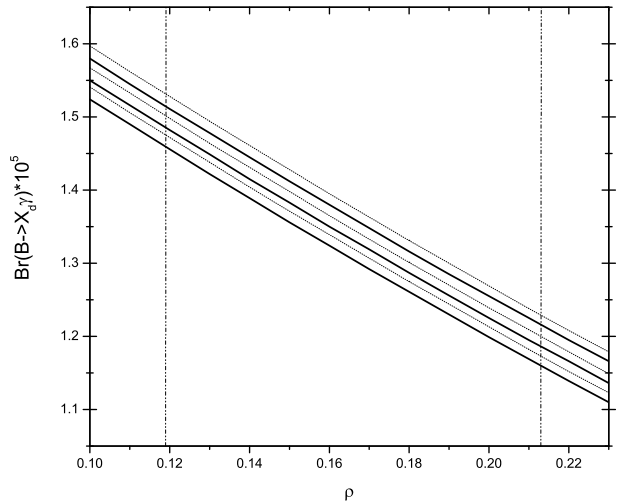}
\end{center}
\caption[]{{\bf{ The dependence of ${\mathcal B}(\bar{B}\to X_d\gamma)$
on the CKM parameters $\rho$ and $\eta$. 
Thick (thin) lines correspond to the branching ratio with matrix elements of QCD penguin operators included
(excluded). The dependence on $\rho$ is shown for three values of $\eta$: $\eta=0.384$ (upper curves),
$\eta=0.356$ (middle curves) and $\eta=0.328$ (lower curves). Vertical lines illustrate the experimental
bounds on the parameter $\rho$.
}}
\label{fig:brandep}}
\end{figure}
As it was pointed out in \cite{Gambino:2001ew}, in the phenomenological analysis
of $\bar{B}\to X_s\gamma$ an important issue is the definition of the mass of charm quark
$m_c$ in the matrix elements of the current-current  operators $O_1, O_2$. The pole mass
have been used in earlier works (e.g. \cite{Greub:1996tg}). However, as it was argued in
\cite{Gambino:2001ew}, the use of the running mass $m_c(\mu)$ with $\mu$ varying between
$m_c$ and $m_b$ is more appropriate. We follow this choice using 
\be\label{zdef}
\frac{m_c}{m_b}=0.23\pm 0.05
\ee
in the matrix elements of the effective operators. Eq. (\ref{zdef}) agrees
with the numbers used in \cite{Hurth:2003dk}. They are obtained
using $m_c(m_c)=(1.25\pm 0.10)\,\, {\rm {GeV}}$ and 
$m_b\equiv m_b^{1S}=(4.69\pm 0.03)\,\,{\rm {GeV}}$ \cite{Hoang:2000fm}.
The latter value for $m_b$ is used also anywhere else in the analysis.

Using these input parameters and the photon energy cutoff $E_0=1.6 {\rm GeV}$
\footnote{
This corresponds to $\delta=0.318$ for $m_b=4.69 {\rm GeV}$.
}
 we obtain for the branching ratios:
\bea\label{brnums}
{\mathcal B}(\bar{B}\to X_s\gamma)&=&\left(3.52^{+0.03}_{-0.16}
(\mu_b)^{+0.00}_{-0.07}(\mu_W)^{+0.22}_{-0.24}(z)^{+0.02}_{-0.02}({\rm CKM})
\pm 0.16 ({\rm param})\right)\times 10^{-4},\\
\label{brnumd}
{\mathcal B}(\bar{B}\to X_d\gamma)&=&\left(1.34^{+0.01}_{-0.08}
(\mu_b)^{+0.00}_{-0.03}(\mu_W)^{+0.13}_{-0.12}(z)^{+0.18}_{-0.17}({\rm CKM})
\pm 0.07 ({\rm param})\right)\times 10^{-5}.
\eea
The central values here correspond to the central values of the input parameters and
 $\mu_W=M_W,\; \mu_b=m_b$. We allow
$\mu_b$ to vary between $m_b/2$ and $2m_b$ and the matching scale $\mu_W$
between $M_W$ and $m_t^{pole}$. The last error sums up the uncertainties due to
$\alpha_s(M_z), m_t^{pole}, {\cal B}(\bar{B}\to X_ce\bar{\nu})^{exp}$ and the factor $C$ 
(see Eq. (\ref{branching})).
The effect of the matrix elements of the QCD penguin operators calculated in this paper
is $-1.0\%$ for $\bar{B}\to X_s\gamma$ and $-1.1\%$ for $\bar{B}\to X_d\gamma$. These corrections
lead to a slight increase of the renormalization scale dependence. Combining all the uncertainties
we get
\bea\label{brnumdf}
{\mathcal B}(\bar{B}\to X_s\gamma)&=&(3.45\pm 0.30)\times 10^{-4},
\\
{\mathcal B}(\bar{B}\to X_d\gamma)&=&(1.27\pm 0.24)\times 10^{-5}
\eea

As we have mentioned in Section 2 in our numerical calculations we have neglected the
long-distance contributions due to presence of intermediate $u$-quark loops. Due to CKM-suppression
these corrections are negligible for $B\to X_s\gamma$ but can be sizeable for $B\to X_d\gamma$. 
This is another source of uncertainty of theoretical prediction for $B\to X_d\gamma$ that can be
roughly estimated to be of order of 10\% (see e.g. \cite{Ali:1998rr} for a more detailed discussion).

The results of our numerical analysis are somewhat lower than those from 
\cite{Gambino:2001ew},\cite{Buras:2002tp}. Note that the central values in Eq. (\ref{brnumd})
will decrease by about 1\% if we use the values from \cite{Gambino:2001ew} for the input parameters.
The numerical discrepancy is explained by the difference in methods used in the analysis.
The main difference is that in \cite{Gambino:2001ew} the charm-quark and top-quark contributions  are split and
each of them is treated separately. 

As we can see from Eq. (\ref{brnumdf}) the uncertainty due to CKM parameters while being
negligible for  $\bar{B}\to X_s\gamma$ is still dominating for
$\bar{B}\to X_d\gamma$. In Fig.\ref{fig:brandep} we illustrate the dependence of 
${\mathcal B}(\bar{B}\to X_d\gamma)$
on $\rho$ and $\eta$.

To conclude, we have presented an independent calculation of the ${\cal O}(\alpha_s)$
corrections to the matrix elements of the QCD penguin operators $O_3,..., O_6$. The effect
of these corrections is about 1\% decrease for the branching ratios. The updated predictions
for the branching ratios are  $(3.45\pm 0.30)\times 10^{-4}$ for $\bar{B}\to X_s\gamma$ and 
$(1.27\pm 0.24)\times 10^{-5}$ for $\bar{B}\to X_d\gamma $.

{\flushleft{
{\large\bf{Acknowledgments}}\\
}}
\vspace{0.5cm}

The work was partially supported by NFSAT-PH 095-02 (CRDF 12050) program. The work H. M. A. was partially 
supported by NATO Grant PST.CLG.978154.
\\

{\flushleft{
\large\bf{Appendix A: Formulae for one-loop building blocks}
}}
\vspace{0.5cm}\noindent
\setcounter{equation}{0}
\renewcommand{\theequation}{A.\arabic{equation}}

In this appendix we will give the complete formulae of the one-loop building blocks
for all different contributions described in section 3.1.
All the calculations are done using the dimensional regularization in $d=4-2\epsilon$
dimensions using the NDR scheme with fully anticommuting $\gamma_5$.
For both diagrams index $\beta$ is to be contracted to the gluon propagator
and $\alpha$ to the photon polarization vector. $r$ and $q$ denote the 4-momenta
of the gluon and photon respectively.
\begin{figure}[htb]
\centerline{
\epsfxsize=4.0in
\epsffile{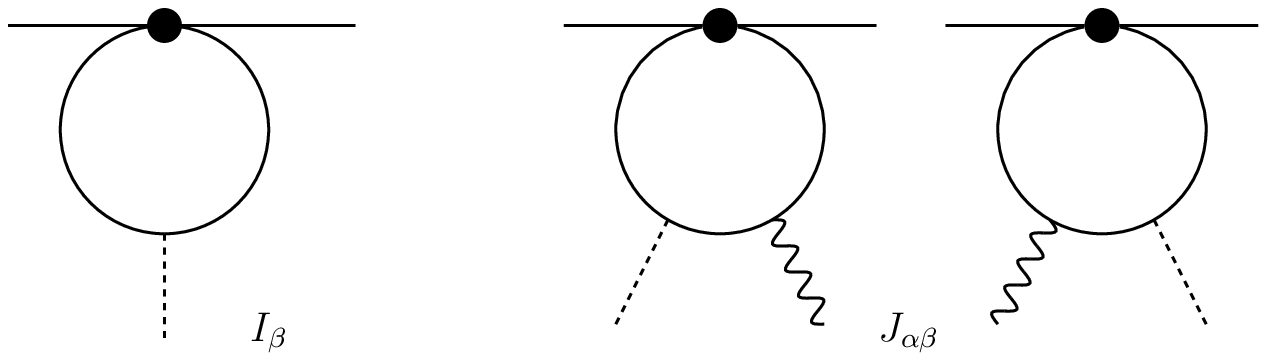}
}
\caption[]{{\bf{ One-loop building blocks used in our calculation.}}
\label{fig:buildingBlocks}}
\end{figure}

For type I insertion the contribution of the $q$-quark loop for the operators
$O_{iL}$ and $O_{iR}$ $(q=s,b;\,\,\,i=3,...,6)$ are respectively 
\bea
I_{\beta(i)}^{q(I)}(L)&\equiv&
  -\bar{C}_{i}^{I}D_{i}^{I}(L)\frac{g_s}{4\pi^2}
  \Gamma(\epsilon)(1-\epsilon)  \mu^{2\epsilon}e^{\gamma_{E}\epsilon}e^{i\pi\epsilon}
  \int\limits_{0}^{1}\frac{dx{[x(1-x)]}^{1-\epsilon}}{\Delta^{\epsilon}}
  (r_{\beta}r\hspace{-0.2cm}/-r^2\gamma_{\beta})L\,\frac{\lambda}{2},\\
I_{\beta(i)}^{q(I)}(R)&\equiv&
  -\bar{C}_{i}^{I}D_{i}^{I}(R)\frac{g_s}{4\pi^2}
  \Gamma(\epsilon)\epsilon \mu^{2\epsilon}e^{\gamma_{E}\epsilon}e^{i\pi\epsilon}
  \int\limits_{0}^{1}\frac{xdx{[x(1-x)]}^{-\epsilon}}{\Delta^{\epsilon}}
  m_q(r\hspace{-0.2cm}/\gamma_{\beta}-\gamma_{\beta}r\hspace{-0.2cm}/)R(q)\,\frac{\lambda}{2},
\eea
where $R(s)=L, R(b)=R$,  
\be
\Delta=r^2-{m_q^2}/{x(1-x)}+i\delta.
\ee
The operator-dependent color coefficients $\bar{C}_i$ and Dirac algebra induced
factors $D_i$ are given by
\begin{eqnarray}
\nonumber
&&\bar{C}_3^{I}=\bar{C}_5^{I}=1,\,\,\,\,\,\bar{C}_4^{I}=\bar{C}_6^{I}=-1/(2N_c);\\
\nonumber
&&D_3^{I}(L,R)=D_4^{I}(L,R)=1,\\
\nonumber
&&D_5^{I}(L)=D_6^{I}(L)=4(4-\epsilon-\epsilon^2),\,\,\,\,
D_5^{I}(R)=D_6^{I}(R)=4(5-3\epsilon-\epsilon^2).
\end{eqnarray}

For the type II insertions $I_{\beta}(R)$ apparently vanishes so we have
$(q=u,d,s,c,b;\,\,\,i=4,6)$
\begin{eqnarray}
\nonumber
I_{\beta(i)}^{q(II)}&\equiv& I_{\beta(i)}^{q(II)}(L)=\\
  &=&-\bar{C}_{i}^{II}D_{i}^{II}\frac{g_s}{4\pi^2}
  \Gamma(\epsilon)\mu^{2\epsilon}e^{\gamma_E\epsilon}e^{i\pi\epsilon}
  \int\limits_{0}^{1}\frac{dx{[x(1-x)]}^{1-\epsilon}}{\Delta^{\epsilon}}
  \left(r_{\beta}r\hspace{-0.2cm}/-r^2\gamma_{\beta}\right)L\,\frac{\lambda}{2}
\end{eqnarray}
where
\begin{eqnarray}
\nonumber
&&\bar{C}_3^{II}=\bar{C}_5^{I}=0,\,\,\,\,\,\bar{C}_4^{II}=\bar{C}_6^{II}=(1/2);\\
\nonumber
&&D_4^{II}=2,\,\,\,\,\,D_6^{II}=4(5-3\epsilon).
\end{eqnarray}

Similarly, for the type I contributions of operators
$O_{iL}$ and $O_{iR}$ for the building block $J_{\alpha\beta}$ we have
\begin{eqnarray}
 \nonumber
J_{\alpha\beta(i)}^{q(I)}(L)&=&A_{i}^{I}B_{i}^{I}(L)\frac{eg_sQ_q}{16\pi^2}\left[
E(\alpha,\beta,r)\Delta i_5+ E(\alpha,\beta,q)\Delta
i_6-E(\beta,r,q)\frac{r_{\alpha}}{qr} \Delta i_{23}\right.\\
&&\left.- E(\alpha,r,q)\frac{r_{\beta}}{qr}\Delta i_{25}
-E(\alpha,r,q)\frac{q_{\beta}}{qr}\Delta i_{26} \right] L\,
\frac{\lambda}{2},\\
\nonumber
J_{\alpha\beta(i)}^{q(I)}(R)&=&A_{i}^{I}m_q\epsilon\frac{eg_sQ_q}{16\pi^2}\Gamma(\epsilon)
\int_SdxdyC^{-1-\epsilon}
\left[8\left(g_{\alpha\beta}(qr)-r_{\alpha}q_{\beta}\right)\right.\\
&& \left.\times\left(dxyB_{i}^{1I}(R)-2B_{i}^{2I}(R)\right)
+2(2+\epsilon)B_{i}^{3I}(R)(r\fnot\gamma_{\beta}\gamma_{\alpha}q\fnot+
q\fnot\gamma_{\alpha}\gamma_{\beta}r\fnot)\right]L\,\frac{\lambda}{2}
\end{eqnarray}
where
\be
E(\alpha,\beta,r)=\frac{1}{2}\left(\gamma_{\alpha}\gamma_{\beta}r\hspace{-0.2cm}/-
r\hspace{-0.2cm}/\gamma_{\beta}\gamma_{\alpha}\right).
\ee
The operator-dependent color coefficients
$A_i$ and Dirac algebra induced factors $B_i$ are given by
\begin{eqnarray}
\nonumber
&&A_3^{I}=A_5^{I}=1,\,\,\,\,\,A_4^{I}=A_6^{I}=-1/(2N_c);\\
\nonumber
&&B_3^{jI}(L,R)=B_4^{jI}(L,R)=1,\,\,\,\,j=1,2,3,\\
\nonumber
&&B_5^{I}(L)=B_6^{I}(L)=4(4-5\epsilon-\epsilon^2),\,\,\,\,
B_5^{1I}(R)=B_6^{1I}(R)=4(1+\epsilon-\epsilon^2),\\
&&B_5^{2I}(R)=B_6^{2I}(R)=4(1-3\epsilon-3\epsilon^2),\,\,\,\,
B_5^{3I}(R)=B_6^{3I}(R)=4(1-7\epsilon-\epsilon^2).
\end{eqnarray}
Finally the quantities $\Delta_i$ are defined as in \cite{Greub:1996tg}
\begin{eqnarray}
\nonumber &&\Delta i_{23}=-\Delta i_{26}=8(qr) \int_S dxdy \left[
x y \epsilon (1+\epsilon) \Gamma(\epsilon) \exp(\gamma_E \epsilon)
\mu^{2\epsilon} C^{-1-\epsilon}\right],\nonumber \\&&\Delta
i_{25}=-8(qr) \int_S dxdy \left[ x(1-x) \epsilon (1+\epsilon)
\Gamma(\epsilon) \exp(\gamma_E \epsilon) \mu^{2\epsilon}
C^{-1-\epsilon}\right],\\\nonumber
&&\Delta i_5=\Delta i_{23},\,\,\, \Delta i_6=\frac{r^2}{(qr)}\Delta i_{25}+\Delta i_{26},
\end{eqnarray}
where $C$ is given by
\begin{eqnarray}
\nonumber && C=m_q^2-2xy(qr)-x(1-x)r^2-i\delta 
\end{eqnarray}
In the case of type II contributions we have
\begin{eqnarray}
 \nonumber
J_{\alpha\beta(i)}^{q(II)}(L)&=&A_{i}^{II}B_{i}^{II}(L)\frac{eg_sQ_q}{16\pi^2}\left[
E(\alpha,\beta,r)\Delta i_5+ E(\alpha,\beta,q)\Delta
i_6-E(\beta,r,q)\frac{r_{\alpha}}{qr} \Delta i_{23}\right.\\
&&\left.- E(\alpha,r,q)\frac{r_{\beta}}{qr}\Delta i_{25}
-E(\alpha,r,q)\frac{q_{\beta}}{qr}\Delta i_{26} \right] L\,
\frac{\lambda}{2},
\eea
where
\bea
\nonumber
B_3^{II}&=&B_4^{II}=B_5^{II}=0,\,\,\,B_6^{II}=-24/(6-d)\\
A_6^{II}&=&1/2.
\nonumber
\eea

{\flushleft{
\large\bf{Appendix B: Formulae for bremsstrahlung corrections}
}}
\vspace{0.5cm}\noindent
\setcounter{equation}{0}
\renewcommand{\theequation}{B.\arabic{equation}}

Here we list the formulae for the bremsstrahlung coefficients $f_{ij}^{ab}$
\footnote{Note that our definitions of quantities $f_{ij}^{cc}$ slightly differ from the standard definitions
of $f_{ij}$ (see e.g.\cite{Chetyrkin:1996vx})}
\footnote{The formula for the coefficient $f_{88}$ is the only place where the mass of the final state quark
has to be retained due to logarithmic divergence. This colinear divergences can be resummed to all
orders in perturbation theory \cite{Kapustin:1995fk}. However, for realistic values of $m_s$ and $m_d$
the contribution of $f_{88}$ is under 1\%, so we will ignore this issue using $m_b/m_s=50$
and $m_b/m_d=1000$ in our analysis.}:
\bea
f_{22}^{cc}&=&\frac{16z}{27}\left[
\delta \int_0^{(1-\delta)/z} dt \; (1-zt) \left| \frac{G(t)}{t} + \frac{1}{2} \right|^2+
\int_{(1-\delta)/z}^{1/z} dt \; (1-zt)^2  \left| \frac{G(t)}{t} + \frac{1}{2} \right|^2 
\right],
\nonumber \\
f_{27}^{cc} &=& -\frac{4 z^2}{9} \left[ 
\delta \int_0^{(1-\delta)/z} dt \left( G(t) + \frac{t}{2} \right) \;+\;
\int_{(1-\delta)/z}^{1/z} dt (1-zt)\left( G(t) + \frac{t}{2} \right) \right],
\nonumber \\
f_{78}^{cc} &=& \frac{4}{9} \left[ {\rm Li}_2(1-\delta) - \frac{\pi^2}{6} - \delta 
\ln \delta + \frac{9}{4} \delta - \frac{1}{4} \delta^2 + \frac{1}{12} \delta^3 \right],
\nonumber\\
f_{77}^{cc} &=&\frac{1}{3}\left[10\delta+\delta^2-\frac{2\delta^3}{3}+\delta(\delta-4)\ln{\delta}\right],
\nonumber\\
f_{88}^{cc} &=& \frac{1}{27} \left[ - 2 \ln \frac{m_b}{m_s} 
                     \left( \delta^2 + 2 \delta + 4 \ln(1-\delta) \right) 
\right. \nonumber \\ && \left.
+4{\rm Li}_2(1-\delta) -\frac{2\pi^2}{3} -\delta(2+\delta)\ln\delta + 8\ln(1-\delta)
-\frac{2}{3} \delta^3 + 3 \delta^2 + 7 \delta \right],
\nonumber\\
f_{22}^{uc} &=& \frac{8z}{27}\left[\delta\int_0^{(1-\delta)/z}dt(1-z
t)\left(\frac{G^\ast(t)}{t}+\frac{1}{2}\right)+\int_{(1-\delta)/z}^{1/z}dt(1-z
t)^2\left(\frac{G^\ast(t)}{t}+\frac{1}{2}\right)\right],
\nonumber\\
f_{27}^{uc}&=&-{\frac {1}{27}}\,\delta\, \left(
3-3\,\delta+{\delta}^{2}
 \right)
 \nonumber\\
 \label{fij}
 f_{22}^{uu}&=&-{\frac{2}{81}}\,\delta\, \left( {\delta}^{2}-3 \right),
\eea
where the function $G(t)$ is defined in Eq. (40) of \cite{Chetyrkin:1996vx}. The other coefficients
can be expressed in terms of those defined above:
\bea
f_{11}^{cc} &=&\frac{1}{36}f_{22}^{cc}, \,\,\, f_{12}^{cc}=-\frac{1}{6}f_{22}^{cc},\,\,\,
f_{28}^{cc}=-\frac{1}{3}f_{27}^{cc},\,\,\, f_{17}^{cc}=-\frac{1}{6}f_{27}^{cc},\,\,\,
f_{18}^{cc}=-\frac{1}{6}f_{28}^{cc},
\nonumber\\
f_{11}^{uc}&=&\frac{1}{36}f_{22}^{uc},\,\,\, f_{12}^{uc}=-\frac{1}{3}f_{22}^{uc},\,\,\,
f_{28}^{uc}=-\frac{1}{3}f_{27}^{uc},\,\,\,f_{17}^{uc}=-\frac{1}{6} f_{28}^{uc},
\,\,\, f_{18}^{uc}=-\frac{1}{6}f_{27}^{uc},
\label{fij1}\\
f_{11}^{uu}&=&\frac{1}{36}f_{22}^{uu},\,\,\,\,
f_{12}^{uu}=-\frac{1}{6}f_{22}^{uu};
\nonumber
\eea
Finally we have
\be
\label{fij2}
f_{ji}=f_{ij}^*
\ee
for the quantities defined in Eqs. (\ref{fij}), (\ref{fij1}). In the bremsstrahlung corrections we have
used the same value $m_c/m_b=0.23\pm 0.05$ as in Eq. (\ref{zdef}) .

\end{document}